\documentclass{iau}
\usepackage{graphicx}

\def\HI{H\,{\sc i}}
\def\barolo{$^{\rm 3D}${\sc barolo}}
\def\apostle{{\sc apostle}}
\def\eagle{{\sc eagle}}
\def\things{{\sc things}}
\def\littlethings{{\sc little~things}}

\title[The APOSTLE simulations] 
{The APOSTLE simulations: Rotation curves derived from synthetic $21$-${\rm cm}$ observations}

\author[Kyle A. Oman]   
{Kyle A. Oman$^1$}

\affiliation{$^1$Department of Physics and Astronomy, University of Victoria, \\
  Victoria, BC V8P 5C2, Canada \\ email: {\tt koman@uvic.ca}
}

\pubyear{2017}
\volume{334}  
\jname{Rediscovering our Galaxy}
\editors{C. Chiappini, I. Minchev, E. Starkenburg, M. Valentini, eds.}
\begin{document}

\maketitle

\begin{abstract}
The \apostle\ cosmological hydrodynamical simulation suite is a collection of twelve regions $\sim5\,{\rm Mpc}$ in diameter, selected to resemble the Local Group of galaxies in terms of kinematics and environment, and re-simulated at high resolution (minimum gas particle mass of $10^4\,{\rm M}_\odot$) using the galaxy formation model and calibration developed for the \eagle\ project. I select a sample of dwarf galaxies ($60<V_{\rm max}/{\rm km}\,{\rm s}^{-1}<120$) from these simulations and construct synthetic spatially- and spectrally-resolved observations of their $21$-${\rm cm}$ emission. Using the \barolo\ tilted-ring modelling tool, I extract rotation curves from the synthetic data cubes. In many cases, non-circular motions present in the gas disc hinder the recovery of a rotation curve which accurately traces the underlying mass distribution; a large central deficit of dark matter, relative to the predictions of cold dark matter N-body simulations, may then be erroneously inferred.
\end{abstract}

\firstsection 
\section{Introduction}

\subsection{The APOSTLE simulation suite}

The \apostle\footnote{A Project Of Simulating The Local Environment.} simulation suite \cite[(Fattahi \etal, 2016a; Sawala \etal, 2016)]{Fattahi2016a, Sawala2016} is a collection of fully cosmological hydrodynamical simulations of $12$ regions each approximately $5\,{\rm Mpc}$ in diameter. These are selected from a large cosmological N-body simulation to resemble the Local Group of galaxies in terms of the properties of the two most massive objects, analogous to the Milky Way (MW) and M~31, in terms of their virial masses, approach and tangential velocity, recession velocity of surrounding objects, and isolation from more massive structures. Each region is re-simulated at multiple resolution levels (see Table~\ref{tab1}) using the modified {\sc P-Gadget3} smoothed-particle hydrodynamics code and calibrated galaxy formation model developed for the \eagle\ project \cite[(specifically, the model is the one denoted `Ref' by Schaye \etal, 2015; see also Crain \etal, 2015)]{Schaye2015,Crain2015}. At present, $5$ of the simulation volumes have been evolved at L1 resolution; all $12$ volumes have been evolved at L2 and L3 resolution.

\begin{table}
  \begin{center}
    \caption{Particle masses and force softening for the three \apostle\ resolution levels, denoted AP-L3 (lowest) to AP-L1 (highest).}
    \label{tab1}
          {\scriptsize
            \begin{tabular}{cccc}
              {\bf Resolution level} & {\bf DM particle$^1$ mass [${\rm M}_\odot$]} & {\bf Gas particle mass [${\rm M}_\odot$]} & {\bf Force softening [${\rm pc}$]}\\
              \hline
              AP-L3 & $7.3\times10^6$ & $1.5\times10^6$ & 711 \\ 
              AP-L2 & $5.9\times10^5$ & $1.3\times10^5$ & 307 \\ 
              AP-L1 & $5.0\times10^4$ & $1.0\times10^4$ & 134 \\ 
            \end{tabular}
          }
  \end{center}
  \vspace{1mm}
  \scriptsize{
    {\it Notes:}\\
    $^1$Particle masses vary by up to a factor of $2$ from volume to volume; values given are indicative.\\
  }
\end{table}

The \eagle\ model is calibrated to reproduce the $z=0.1$ galaxy stellar mass function (GSMF) and galaxy size distribution across the range $10^8<M_\star/{\rm M}_\odot<10^{11}$. The \apostle\ suite demonstrates that the same model, without recalibration, reproduces on average the Local Group GSMF down to $M_\star=10^5\,{\rm M}_\odot$ \cite[(Sawala \etal, 2016)]{Sawala2016}, and the size distribution down to $M_\star=3\times10^6\,{\rm M}_\odot$ \cite[(Campbell \etal, 2017)]{Campbell2017}. The \apostle\ MW and M~31 analogs do not suffer from the `too-big-to-fail' problem, defined by \cite[Boylan-Kolchin \etal\ (2011)]{BoylanKolchin2011}: their satellite galaxies occupy dark matter halos consistent with the observed velocity dispersions of dwarf spheroidal galaxies in the Local Group \cite[(Sawala \etal, 2016; Fattahi \etal, 2016b)]{Sawala2016,Fattahi2016b}.

The simulation suite has been used to assist studies of stellar disc dynamics (Yozin \etal\ in preparation), the structure of gaseous discs \cite[(Ben\'{i}tez-Llambay \etal, 2017)]{BenitezLlambay2017}, the assembly of stellar halos \cite[(Starkenburg \etal, 2017; Oman \etal, 2017b)]{Starkenburg2017,Oman2017b}, the baryonic Tully-Fisher relation \cite[(Oman \etal, 2016; Sales \etal, 2017)]{Oman2016,Sales2017}, the properties of low-mass, starless dark matter haloes \cite[(Ben\'{i}tez-Llambay \etal, 2016; Sawala \etal, 2017)]{BenitezLlambay2016,Sawala2017}, the tidal stripping of satellite galaxies \cite[(Fattahi \etal, 2017; Wang \etal, 2017)]{Fattahi2017,Wang2017}, the mass discrepancy -- acceleration relation \cite[(Ludlow \etal, 2017; Navarro \etal, 2017)]{Ludlow2017,Navarro2017}, and the dark matter cusp-core problem \cite[(Oman \etal, 2015; 2017a; Genina \etal, 2017)]{Oman2015,Oman2017a,Genina2017}.

\subsection{The dark matter cusp-core problem}

N-body simulations generically predict that cold dark matter halos have radial density profiles which rise steeply toward the halo centre, with a logarithmic slope of $\approx -1$ \cite[(Navarro \etal, 1996a; 1997)]{Navarro1996a,Navarro1997}, termed a cusp. The density profiles for some galaxies, as inferred from their rotation curves or other dynamical mass estimators, however, imply an approximately constant central dark matter density (slope $\approx 0$) within some `core' region \cite[(Moore 1994; Flores \& Primack 1994; and see de Blok 2010 for a review)]{Moore1994,FloresPrimack1994,deBlok2010}. The same discrepancy can also be cast as a central deficit of mass, relative to the prediction from N-body simulations, as in \cite[Oman \etal\ (2015)]{Oman2015}. An important constraint on any proposed solution to this problem is the diversity in the rotation curves of galaxies at fixed maximum rotation velocity $V_{\rm max}$, especially those of dwarfs of $30\lesssim V_{\rm max}/{\rm km}\,{\rm s}^{-1}\lesssim 100$. Any mechanism proposed to create cores -- or the appearance of a core -- must do so in some galaxies and not in others, and to different extent in different galaxies.

Currently viable resolutions of the cusp-core problem can be summarized as falling into several categories (or combinations thereof):
\begin{itemize}
\item The dark matter distribution, presumed initially cuspy, may be modified by gravitational coupling to violent motions of gas within the galaxy \cite[(Navarro \etal, 1996b; Read \& Gilmore, 2005; and for a review see Pontzen \& Governato, 2014)]{Navarro1996b,ReadGilmore2005,PontzenGovernato2014}.
\item If the dark matter physics differs from that of generic cold dark matter, for instance if the dark matter is warm \cite[(Bode \etal, 2001; Lovell \etal, 2012)]{Bode2001,Lovell2012} or self-interacting \cite[(Spergel \& Steinhardt, 2000; Creasey \etal, 2017)]{SpergelSteinhardt2000,Creasey2017}, the density profile may be modified.
\item Systematic effects and errors in the modelling of observed galaxies could cause the appearance of a core in a system hosting a dark matter cusp \cite[(Rhee \etal, 2004; Valenzuela \etal, 2007; Kuzio de Naray \& Kaufmann, 2011; Read \etal, 2016; Pineda \etal, 2017; Oman \etal, 2017a; and references therein)]{Rhee2004,Valenzuela2007,KuziodeNarayKaufmann2011,Read2016,Pineda2017,Oman2017a}.
\end{itemize}

\section{Method}

\subsection{Synthetic observations}\label{synthobs}

I select the $33$ galaxies from the \apostle\ L1 resolution simulations which lie in the interval $60 < V_{\rm max}/{\rm km}\,{\rm s}^{-1} < 120$. The upper bound ensures that the galaxies are `dark matter-dominated' in the sense that when the gravitational force is decomposed into contributions from dark matter and stars and gas, the component due to the dark matter is dominant, on average, at every radius. The lower bound ensures that corrections to the rotation curve to account for `pressure support' \cite[(for a detailed discussion of such corrections see Valenzuela \etal, 2007; Pineda \etal, 2017)]{Valenzuela2007,Pineda2017} are usually small.

The \HI\ distribution in each galaxy is computed from the particle positions, chemical compositions, temperatures and densities following the prescription of \cite[Rahmati \etal\ (2013)]{Rahmati2013} for self-shielding from the metagalactic ionizing background radiation, with a correction for the molecular gas fraction following \cite[Blitz \& Rosolowsky (2006)]{BlitzRosolowsky2006}. Each galaxy is then `observed' to produce a synthetic data cube, mimicking where possible the characteristics of the \littlethings\ \HI\ survey \cite[(Hunter \etal, 2012)]{Hunter2012}. The galaxies are placed at a nominal distance of $3\,{\rm Mpc}$ and an inclination of $60^\circ$. Another angle, $\Phi$, corresponding to an azimuthal rotation of the galaxy, needs to be chosen to specify a unique line of sight -- this angle is randomly selected. The \HI\ distribution is mapped onto a grid with a pixel spacing of $3\,{\rm arcsec}$ and convolved with a $6\,{\rm arcsec}$ gaussian beam. In the velocity direction, each particle contributes a gaussian line profile of $7\,{\rm km}\,{\rm s}^{-1}$ fixed width, and amplitude proportional to the \HI\ mass of the particle. In each pixel the velocity information is sampled in channels of $4\,{\rm km}\,{\rm s}^{-1}$ width.

\begin{figure}[b]
  \begin{center}
    \includegraphics[width=3.2in]{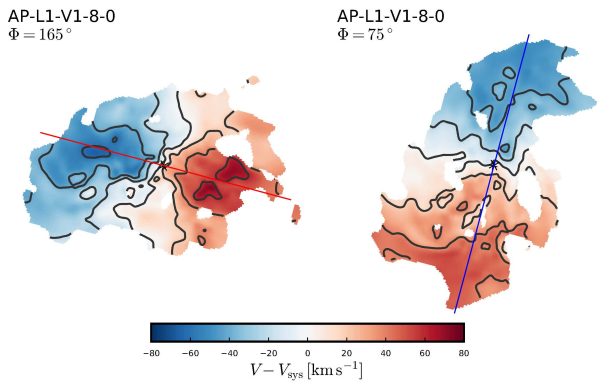} 
    \caption{`Synthetically observed' \HI\ intensity-weighted mean velocity field for the simulated dwarf galaxy AP-L1-V1-8-0 viewed from two directions separated by $90^\circ$, at fixed inclination $i=60^\circ$. The straight lines indicate the kinematic major axis of each projection. (Adapted from \cite[Oman \etal\ 2017]{Oman2017a}, fig.~8.)}
    \label{fig1}
  \end{center}
\end{figure}

Fig.~\ref{fig1} is a visualization of one of the simulated galaxies, AP-L1-V1-8-0\footnote{AP-[resolution level]-[volume number]-[friends-of-friends group number]-[subgroup number].} seen from two viewing angles separated by $90^\circ$, at a fixed inclination of $60^\circ$. Note that, for illustrative purposes, the viewing angles have been carefully chosen, rather than randomized as described above. The maps show the intensity-weighted mean velocity in each pixel, and the straight lines indicate the kinematic major axis of each projection. Galaxies at this mass scale have substantial azimuthal asymmetries, reflected here as differences between the velocity fields after an azimuthal rotation.

\subsection{Rotation curve modelling}

Using the \barolo\ tilted-ring modelling software \cite[(Di Teodoro \& Fraternali, 2015)]{DiTeodoroFraternali2015}, a rotation curve is extracted from each synthetic data cube. \barolo\ does not model the velocity field, but rather constructs a full model data cube whose residual with respect to the observed data cube is minimized via an iterative variation of the parameters of each `ring' in the model. The configuration used is as in \cite[Iorio \etal\ (2017)]{Iorio2017}, except where parameters need to be set on a per-galaxy basis (e.g. centroid, size, inclination, position angle). The centroid is fixed to the peak of the simulated stellar particle distribution as seen along the `line of sight'. The inclination and position angles are set to the known values used to construct the synthetic observations, but allowed to vary in the fitting process by up to $15^\circ$ and $20^\circ$, respectively.

The rotation curves are corrected for pressure support as:
\begin{equation}
  V_{\rm circ}^2 = V_{\rm rot}^2 - \sigma^2\frac{{\rm d}\log(\Sigma_{\rm HI}\sigma^2)}{{\rm d}\log R}
\end{equation}
where $V_{\rm rot}$ is the rotation curve as recovered by \barolo, $V_{\rm circ}$ is the corrected rotation curve, $\sigma$ is the \HI\ velocity dispersion and $\Sigma_{\rm HI}$ is the \HI\ surface density -- all these quantities are measured directly from the synthetic data cubes.

\section{Results}

\begin{figure}[b]
  \begin{center}
    \includegraphics[width=3.2in]{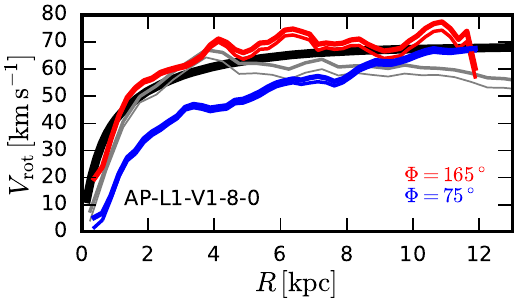} 
    \caption{The thick black curve is the circular velocity profile of simulated galaxy AP-L1-V1-8-0 (the same object used in Fig.~\ref{fig1}). The thin and thick gray lines indicate the rotation speed of \HI\ gas in this galaxy as a function of radius, before and after correction for pressure support, respectively. The thin red and blue curves are the rotation curves obtained from a \barolo\ modelling of the synthetic \HI\ datacubes corresponding to the two orientations for which the velocity fields are shown in Fig.~\ref{fig1}. The application of a correction for pressure support yields the thick red and blue curves. (Adapted from \cite[Oman \etal\ 2017]{Oman2017a}, fig.~5.)}
    \label{fig2}
  \end{center}
\end{figure}

The rotation curves for AP-L1-V1-8-0 corresponding to the two orientations shown in Fig.~\ref{fig1} are shown in Fig.~\ref{fig2} as coloured lines; the thicker lines are the pressure support-corrected versions of the rotation curves shown with thin lines. The gray lines show the rotation speed as a function of radius for the simulated \HI\ gas as measured directly from the simulation output, i.e. without the added handicaps of projection, limited `observational resolution', etc. Finally, the thick black line shows the circular velocity curve for this galaxy, computed as $V_{\rm circ}=\sqrt{GM(<R)/R}$, where $M(<R)$ is the mass enclosed within radius $R$, and $G$ is the gravitational constant. The dark matter is the dominant contributor to $M(<R)$ at all radii for all simulated galaxies in the sample. In the orientation labelled $\Phi=165^\circ$, the `observed' rotation curve overestimates somewhat the rotation speed of the gas in the galaxy; in the $\Phi=75^\circ$ orientation, the rotation curve is \emph{severely} underestimated.

The leading reason for the difference between the two rotation curves in Fig.~\ref{fig2} is the presence of a non-circular, bisymmetric flow pattern superimposed on the rotation. This imposes a velocity modulation which has two peaks and two troughs along an azimuthal loop around the galaxy. The rotation curve is driven up when gas rotating faster than the average at that radius falls on the major axis of the galaxy in projection, and down when gas rotating more slowly falls there. Such features are common in the simulated sample and lead to a wide diversity of rotation curve shapes, this despite the fact that all retain their dark matter cusp and are dark matter dominated -- if their rotation curves faithfully traced their circular velocity curves the shapes would all be self-similar.

Though such bisymmetric flow patterns are easily discerned based on the simulation particle properties, their signature in the synthetic observations is more subtle. In some cases where the rotation curve is severely underestimated, I have been unable to identify any clear indication that anything might be amiss; were these real galaxies, I might erroneously come to the conclusion that they host dark matter cores.

Considering the sample of $33$ simulated galaxies as a whole, when each is modelled with a random orientation (at fixed inclination $i=60^\circ$, see Sec.~\ref{synthobs}), the width of the rotation curve shape distribution, parametrized by the rotation speed at $2\,{\rm kpc}$ and $V_{\rm max}$ \cite[(Oman \etal, 2015)]{Oman2015}, is comparable to that seen in real galaxies \cite[(Oman \etal, 2017a)]{Oman2017a}.

\section{Conclusions}

The ability of the scenario outlined above to constitute a convincing resolution of the cusp-core problem hinges on the `realism' of the simulated galaxies. The data cubes constructed for the \apostle\ dwarfs compare favourably with those from the \things\ \cite[(Walter \etal, 2008)]{Walter2008} and \littlethings\ \cite[(Hunter \etal, 2012)]{Hunter2012} surveys according to several metrics \cite[(Oman \etal, 2017a)]{Oman2017a}, but there are some differences even at this rudimentary level of comparison: the \apostle\ galaxies have somewhat higher velocity dispersions on average, and perhaps also thicker \HI\ discs. Nevertheless, there is some indication that bisymmetric non-circular flow patterns sufficiently strong to cause substantial changes to their rotation curves may be present in the DDO~47 and DDO~87 galaxies \cite[(Oman \etal, 2017a)]{Oman2017a}: the residuals after subtraction of a tilted-ring model from their velocity fields reveal a `trefoil' shape characteristic of a bisymmetric distortion seen in projection. I urge continued strong caution when interpreting the rotation curves of dwarf galaxies, especially in the central regions which are often poorly sampled and in which the potential impact of errors due to non-circular motions is strongest.

\end{document}